\documentclass[twocolumn,showpacs,preprintnumbers,amsmath,amssymb]{revtex4}
\usepackage{graphicx,subfigure}% Include figure files
\usepackage{dcolumn}% Align table columns on decimal point
\usepackage{bm}% bold math

\begin{document}

\preprint{APS/123-QED}

\title{Oceanic Ambient Noise as a Background to Acoustic Neutrino Detection}% Force line breaks with \\

\author{Naoko Kurahashi}
 \email{naokok@stanford.edu}
\author{Giorgio Gratta}%
\affiliation{%
Departments of Physics and Applied Physics, Stanford University, Stanford, CA 94305}%Line breaks may be forced with \\ here, too

\date{\today}% It is always \today, today,

\begin{abstract}
Ambient noise measured in the deep ocean is studied
in the context of a search for signals from ultra-high energy cosmic ray neutrinos.
The spectral shape of the noise at the relevant high frequencies is found to be very stable for an extensive data set collected over several months from 49 hydrophones mounted near the bottom of the ocean at $\sim$1600~m depth. The slopes of the ambient noise spectra above 15~kHz are found to roll-off faster than the -6 dB/octave seen in Knudsen spectra.  A model attributing the source to an uniform distribution of surface noise that includes frequency-dependent absorption at large depth is found to fit the data well up to 25~kHz. This depth dependent model should therefore be used in analysis methods of acoustic neutrino pulse detection that require the expected noise spectra.

\end{abstract}

\maketitle

\section{INTRODUCTION}

The study of high frequency acoustic transients in water was first proposed by Askaryan~\cite{Askaryan} as a possible method for the detection of the highest energy cosmic radiation.   In the last few years feasibility studies have been initiated for acoustic ultra-high-energy neutrino detectors in large natural bodies of water, ice and salt.   This technique would present a number of advantages, among them the possibility of building very large arrays (thousands of km$^2$) with sparse hydrophones, owing to the large attenuation length ($\sim$1~dB/km\cite{Fisher}) of sounds at the appropriate frequencies.   This is essential because of the expected extremely low neutrino flux ($\ll$~1~km$^{-2}$~yr$^{-1}$) at the energies of interest (E$>$10$^{18}$~eV).    Because of the small interaction cross section of neutrinos, sound has to be detected in rather deep media, providing a larger probability of interaction per unit area.    The acoustic radiation is produced from the volume expansion caused by the heating of the medium where the neutrino interacts and stops.   Theoretical models~\cite{LearnedTheory} and experimental measurements \cite{Learned} show that the signature of this kind of event is a single bipolar pulse, with energy concentrated around 10~kHz.   Substantial data processing of pulse shapes and multi-phone correlations are required to extract these signals from the ambient noise and from transients produced by specific sources.

It is broadly accepted that underwater ambient noise in the range of 1 to 25~kHz is generated at the sea surface~\cite{Urick}. However, starting with the empirical study of Knudsen et al.\cite{Knudsen}, the extensive data supporting this hypothesis has been collected only in shallow water or over a limited frequency band (or both). The variability of ambient noise with depth is also measured at one or a few frequencies and does not provide the broad range spectral shape. It is unclear as to whether some level of bulk emission (apart from
molecular agitation) might be required to reproduce the noise behavior at high frequencies and large depths. As early as 1962, when attempting to model ambient noise with wave action, Marsh noted that a depth dependent feature should be included~\cite{Marsh}. Short~\cite{Short}, following Urick's treatment of deep water ambient noise, developed a mathematical treatment for the depth and frequency dependent features. These results have not been verified on large data sets and broad frequency bands.   The effort to study acoustic neutrino detection naturally produces very large, wide-band data sets, collected at large depths, making an extensive analysis of ambient noise in this regime possible.

\section{THE SAUND II EXPERIMENT}

The second phase of the "Study of Acoustic Ultra-high energy Neutrino Detection" (SAUND~II) employs a large hydrophone array in the US Navy's Atlantic Undersea Test and Evaluation Center (AUTEC) located at the Tongue of the Ocean (TOTO) in the Bahamas.    SAUND~II is currently the largest test for the feasibility of acoustic ultra-high-energy neutrino detection. This program follows a general study of the expected performance~\cite{Lehtinen}, and a first experimental phase (SAUND~I) using seven hydrophones at the same location~\cite{Vandenbroucke}.    SAUND~II uses 49 hydrophones that are digitized in the water, with data transmitted to shore over optical fibers.    The array spans an area of $\sim 20$~km $\times$ $50$~km with spacing of 3~to~5~km.   Hydrophones are mounted 5.2~m above the ocean floor, at depths between 1340 and 1880~m and are omnidirectional with a flat response ($<$5~dB) from 50~Hz to 15~kHz.   Between 15~kHz and the system cutoff of 40~kHz there is a directional dependence in the response, with up to 22~dB difference between the vertical and horizontal directions at 40 kHz.   The gains of the 49 channels coincide to within 1~dB. Analog signals are regenerated on the shore station from the digital data (for backward compatibility) and fed to the SAUND~II data acquisition system that re-digitizes them at 156~kHz.    Since low frequencies are not relevant for SAUND~II, a high-pass RC filter is applied to the analog data with a 3~dB point at 100~Hz.    A real time analysis program running on seven computers records candidate neutrino events as well as other data of interest. This includes a power spectral density (PSD) obtained by integrating 6.56~ms of data, sampled at 6.4~$mu$s, every 5~s for each hydrophone continuously while the SAUND~II system is on.   The results presented here are an analysis of this PSD data, taken for the purpose of understanding the ambient noise background.   By agreement with the US Navy, the SAUND II data acquisition system records data only when the array is not otherwise used. From July 2006 to September 2007, the system has been running under these conditions stably for a total integrated time of $\sim$150~days.

\section{AMBIENT NOISE DATA}

\begin{figure}
\includegraphics[scale=0.425]{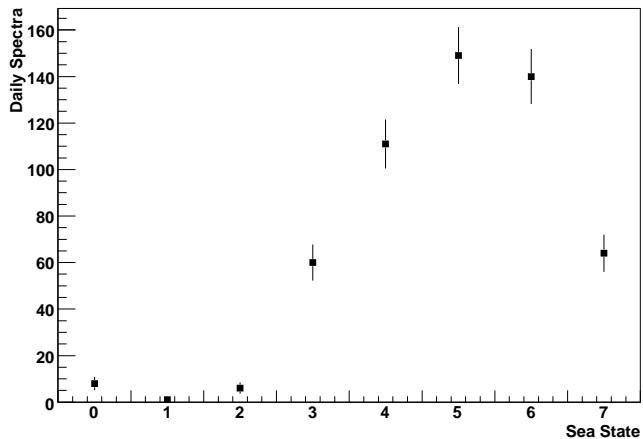}
\caption{Histogram of daily averaged sea state calculated from 539 noise spectra taken by 49 hydrophones over 11 24-hour periods ranging from July to November 2006. The sea state 0 bin includes quiet periods calculated to be under the average sea state 0 noise level. Error bars indicate the magnitude of the statistical fluctuations.}
\label{fig:sea_state}
\end{figure}

Ambient noise analyses are performed on a subset of the data, consisting of 11 data periods randomly spread between July to November 2006.  Each data period consists of 24 hours of continuous data-taking. In order to average-out intermittent sources that are of transient nature, such as ship traffic and fish feeding, all 17280 PSDs in a day are averaged to produce one spectrum for each hydrophone and every data period. For frequencies up to 15~kHz, the resulting 539 spectra all follow the expected $f^{-5/3}$ Knudsen shape.  Therefore, the 1~kHz - 15~kHz region is used to unfold the sea state conditions.  The ambient noise spectrum at sea state zero, in units of dB re $\mu$Pa$^{2}$/Hz, is approximated from the average spectra presented in Urick \cite{Urick} as
\begin{equation}
P(f) = 10\log(f^{-5/3}) + 94.5
\label{eq:seastatezero}
\end{equation}
for the spectral region of 1~kHz to 40~kHz. Since the dependence of the overall noise level on sea state does not influence the $f^{-5/3}$ falloff, noise levels for higher sea states can be found by adding to Eq.~(\ref{eq:seastatezero}), $P_{ss} = 30\log(n_s + 1)$ where $n_s$ represents the sea state \cite{Short}. The resulting analytical form becomes
\begin{equation}
P(f, n_s) = 10\log(f^{-5/3}) + 94.5 + P_{ss}(n_s)
\label{eq:seastate}
\end{equation}
with a continuous variable $n_s$ in $P_{ss}$ as the only free parameter to fit each PSD.  Data are then fit to Eq.~(\ref{eq:seastate}) using the least square method and equally weighing all frequencies from 1~kHz to 15~kHz. The discretized value $n_s$ is used to produce Fig.~\ref{fig:sea_state}. The fact that July to November encompasses hurricane season in the Caribbean, and hence volatile sea conditions, explains the relatively high daily averaged sea states measured.

\begin{figure}
\subfigure{
\includegraphics[scale=0.42]{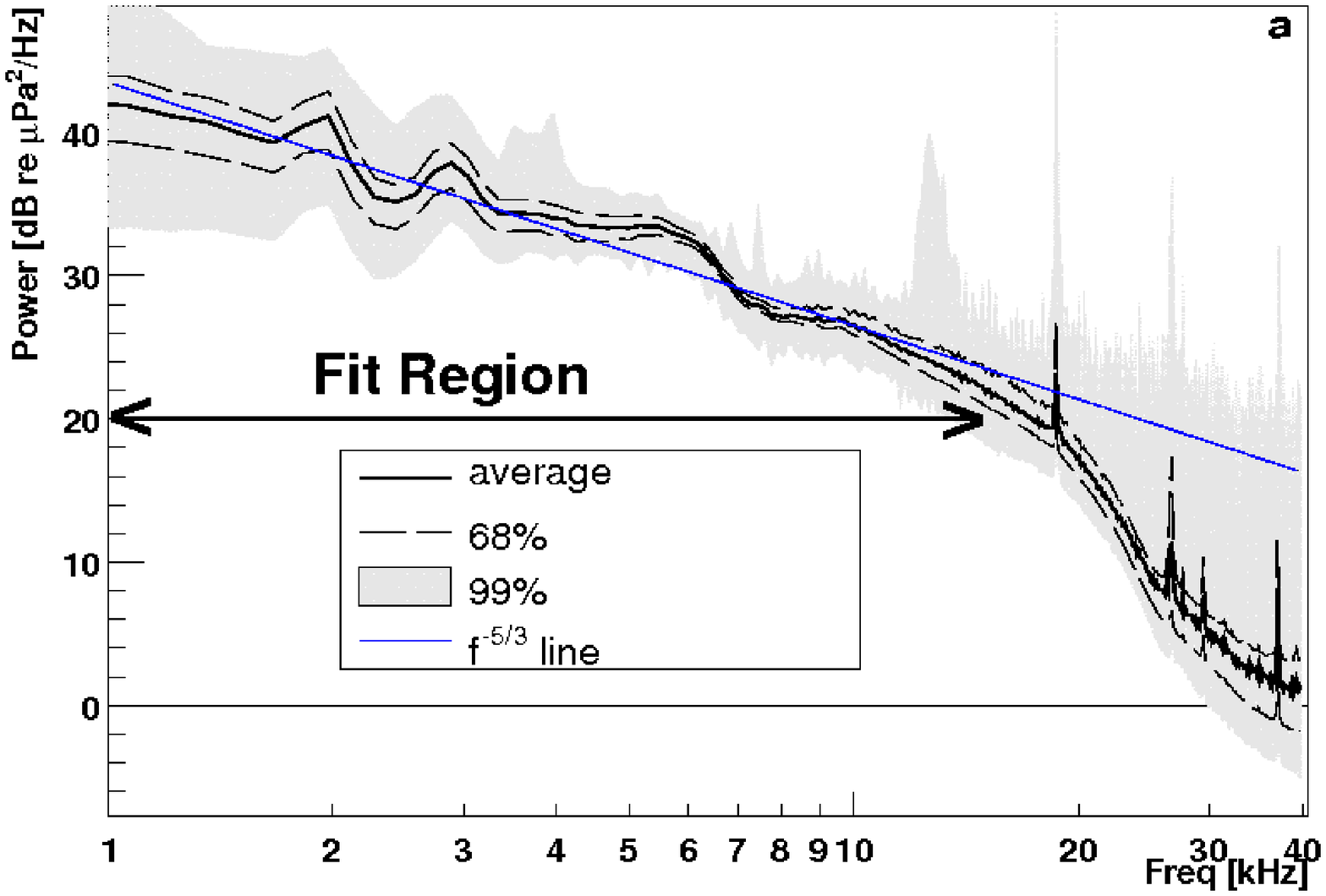}
\label{fig:spreadDaily}
}
\subfigure{
\includegraphics[scale=0.42]{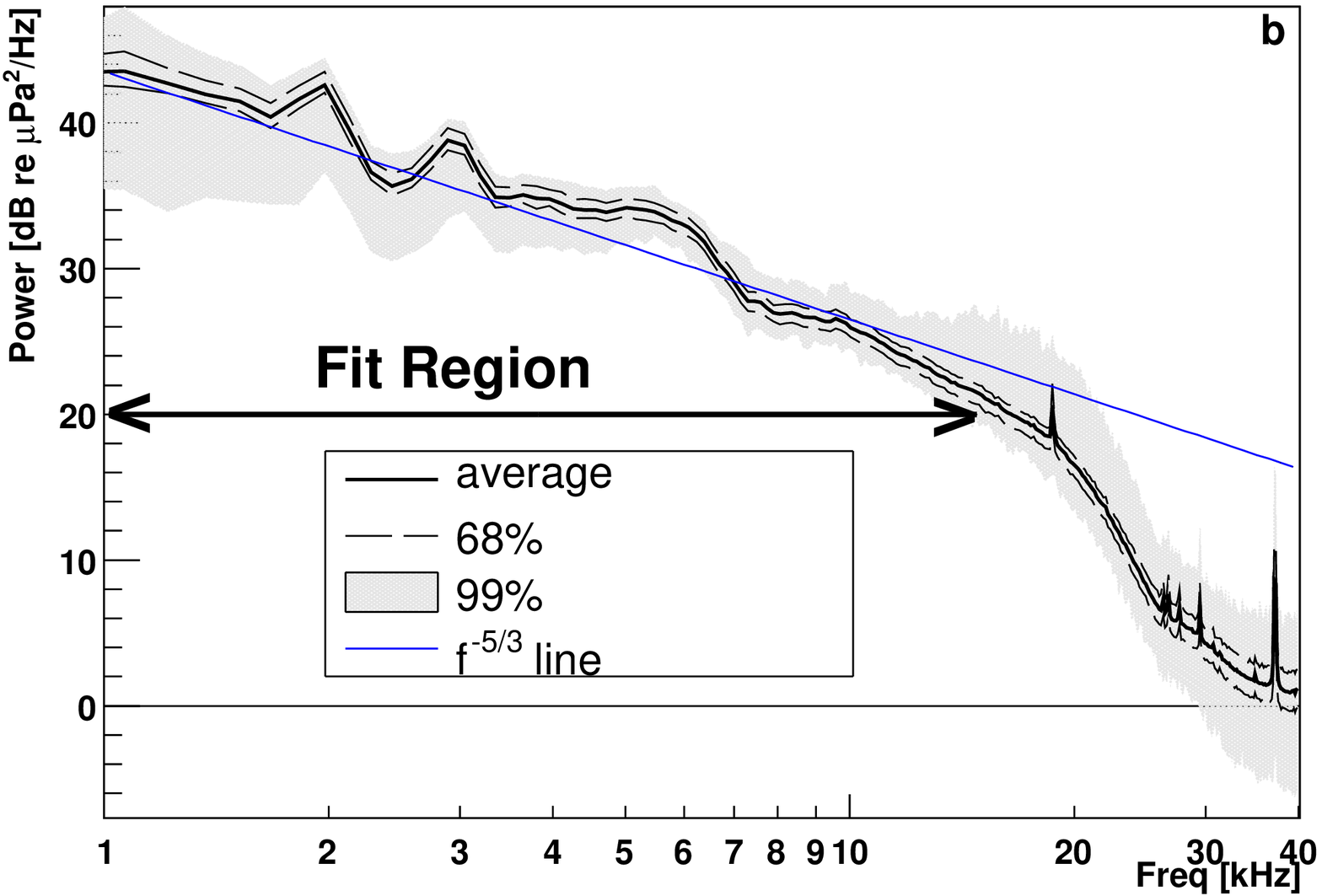}
\label{fig:spreadMinutely}
}
\caption{$P_{ss}$-subtracted power spectral density. The average of the daily (5~min) averaged spectra over 11 days (24~hours) and 49 (1) hydrophones is shown in a (b), along with the envelopes containing 99\% and 68\% of 539 (287) spectra.}
\label{fig:specSpread}
\end{figure}

After the fitted sea state quantity $P_{ss}$ is subtracted from each spectrum, the resulting 539 full bandwidth ambient noise spectra exhibit remarkably similar shapes, even beyond the 1~kHz to 15~kHz range where the $f^{-5/3}$ power law no longer describes the slopes. Fig.~\ref{fig:specSpread}a shows the spread in the data by plotting the envelopes containing 99\% and 68\% of the 539 spectra and their average after "sea state subtraction." It is apparent that while sea state conditions change drastically, the underlying ambient noise spectral shape, even at high frequencies, remains consistent. This is significant considering the different locations and seasons included in the analysis.

In order to check that the ambient noise spectra have shapes that are consistent over even shorter time scales, Fig.~\ref{fig:specSpread}b plots the spread of the "sea state subtracted" spectra averaged over 5 min intervals measured on October 2nd 2006 at one hydrophone. Again the spread is presented in a manner similar to Fig.~\ref{fig:specSpread}a, plotting the envelope containing 99\% and 68\% of the 287 spectra along with the average. The smooth features of the spectrum are stable. The extremely small spread in these spectra disfavors the possibility of intermittent sources contributing to daily averaged "sea state subtracted" spectra over the broad bands analyzed here. This suggests that if noise sources other than those causing surface noise give an important contribution, they must be continuous in time. Very few sources are identified as having a steady all-day impact. Thermal noise is expected to become important at higher frequencies than measured here, while seismic background of continuous disturbance occurs far below 1~kHz \cite{Wenz}. Noise produced by marine biology is expected to have transient and narrow band characteristics. One well known exception is snapping shrimp beds, known for their consistent sound production underwater \cite{Everest}. However, there is no evidence that shrimp beds exist at these depths, and the similarity of the spectra measured over an area of $\sim$1000 km$^{2}$ disfavours this possibility.  Therefore, this analysis proceeds to directly test the hypothesis that surface generated noise is the only dominant broad band source contributing to ambient noise at these depths and frequencies.

\section{CORRELATION WITH WIND}

\begin{figure} 
\includegraphics[scale=0.425]{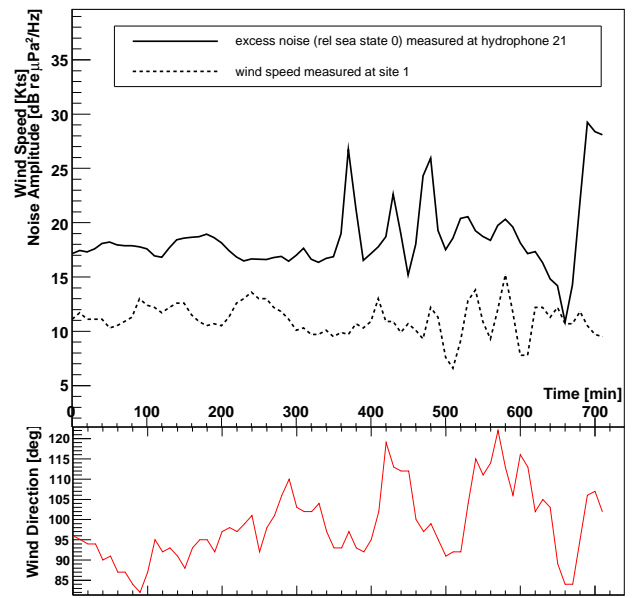} 
\caption{Wind speed and measured offset from the sea state zero, $P_{ss}$, as functions of time starting July 3rd 2006 at 0124~UTC (top panel). Wind direction is also plotted in degrees North (bottom panel). Wind measurements are taken at a site $\sim$21~km away in the range from the hydrophone used here.} 
\label{fig:wind} 
\end{figure}

\begin{figure} 
\includegraphics[scale=0.425]{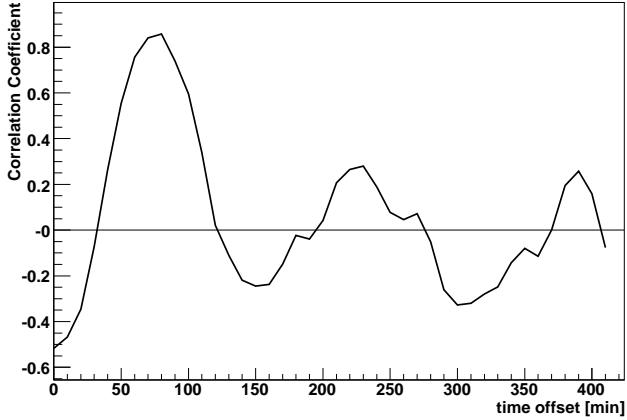}
\caption{Correlation coefficient between the first 300 minutes of noise data from Fig.~\ref{fig:wind} and the equivalent period of wind speed data starting at different time offsets (horizontal axis). The main maximum at a time offset of $\sim$ 80~min is followed by substantially less significant maxima due to the spikes of the quasi-periodic wind at times between 500 and 600 min.} 
\label{fig:correlation} 
\end{figure}

To verify that the overall noise level due to the sea state can indeed be expressed by Eq.~\ref{eq:seastate}, measured wind speeds are compared to the fitted value $P_{ss}$ measured at the hydrophone closest to a shore anemometer station.    Wind speeds, logged every 10 minutes at a site on Andros Island at a height of $\sim$15 m, 21~km away from the vertical of the hydrophone at a bearing of $110^\circ$, are shown in Fig.~\ref{fig:wind}.   In the Figure the direction of the wind and the quantity $P_{ss}$ averaged every 10~min are also shown.   The particular data period chosen for this study encompasses a rare occasion with a rather stable wind direction ($\simeq100^\circ$) over a long period of time (12-hours).  The Pearson product-moment correlation coefficient is computed between the first 300~min of $P_{ss}$ data and an equivalent duration of wind data, applying different time offsets between the two. The correlation coefficient is shown as a function of the time offset in Fig.~\ref{fig:correlation}.    The correlation coefficient reaches a maximum of 0.85 at a time offset of $\simeq$80~min.    This is consistent with the delay expected from the distance, approximate wind speed and direction.  In order to understand the statistical significance of this result, various 12-hour periods of wind data are randomly chosen from the 2006 year, and the same analysis is repeated using the original $P_{ss}$ time series from Fig.~\ref{fig:wind}.  For each 12-hour period, a plot similar to Fig.~\ref{fig:correlation} is produced, and the maximum correlation coefficient within the 0 to 420~min offset is chosen. Of 100 such 12-hour wind data periods, only once a correlation coefficient higher than 0.85 was observed. Because the wind direction is usually quite variable over the travel time between the hydrophone location and the anemometer, conditions like those selected in Fig.~\ref{fig:wind} are rarely found.    In addition there is no guarantee that the wind front stays coherent over the $\sim$1 hour travel to the anemometer station.   Despite these limitations, however, it appears that a clear correlation is observed between the quantity $P_{ss}$ and the wind strength.

\section{DEPTH DEPENDENT AMBIENT NOISE SPECTRUM}

\begin{figure}[htb!!!!!!!!!!!!!!!!!!!!!!!!!!!!!!!!!!!!]
{
\includegraphics[scale=0.425]{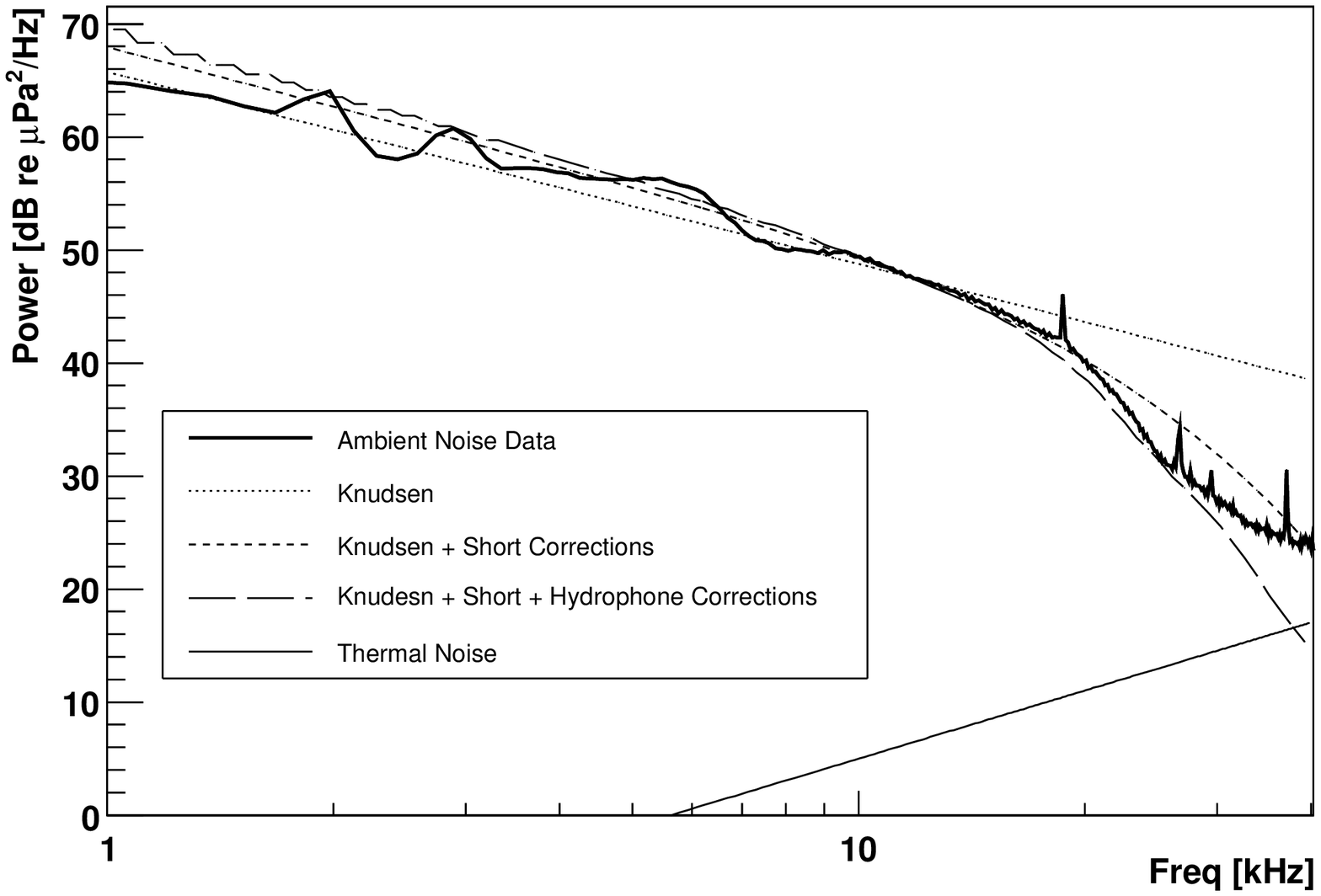}
\label{4a}
}
{
\includegraphics[scale=0.425]{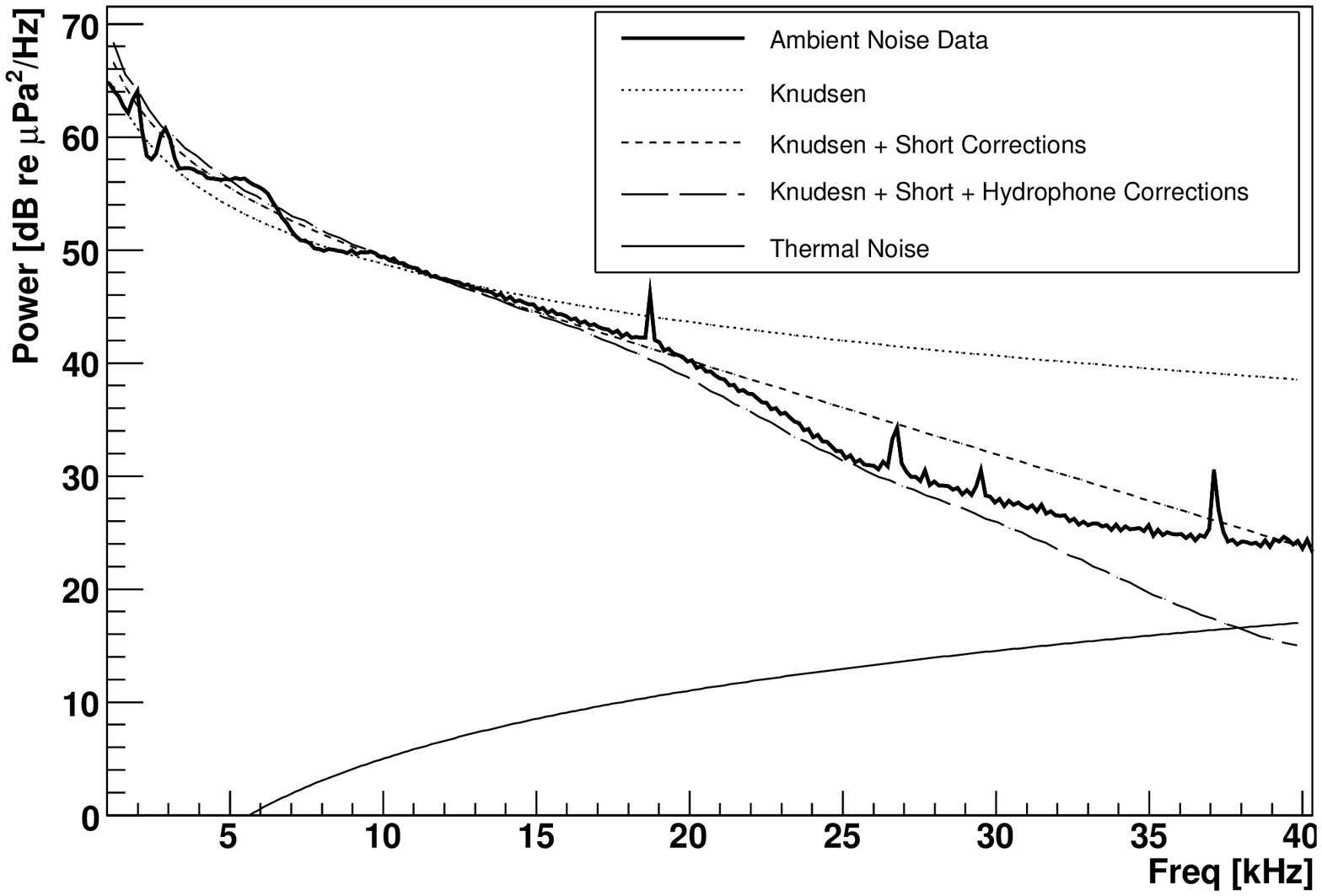}
\label{4b}
} 
\caption{The average ambient noise measured, plotted in log (above) and linear (below) frequency scale with 3 theoretical curves of Knudsen spectra, as discussed in text. The theoretical curve for thermal noise\cite{Urick} is also plotted. The curves labeled "Short corrections" in the legend include frequency dependent attenuation parametrized by Fisher and Simmons \cite{Fisher}.}\label{fig:full_bandwidth_spectrum}
\end{figure}

Shown in Fig.~\ref{fig:full_bandwidth_spectrum} is the average of 539 ambient noise spectra measured without subtracting the increased noise level at different sea states. The expected Knudsen form of Eq.~(\ref{eq:seastate}) is also plotted with a $P_{ss}$ value that fits the average best between 1~kHz and 15~kHz. This analytical spectrum reproduces the data well in this range.   Above this frequency, however, it substantially overestimates the noise.    Next, a depth-dependent modification developed by Short~\cite{Short} is used to modify Eq.~(\ref{eq:seastate}) to account for the large depth. Short's model considers the frequency dependent absorption that becomes significant when hydrophones are placed in deep waters, and thus, the distance from the surface where the noise originates becomes large. This modification is particularly important at high frequencies as absorption becomes substantial.   The effective noise intensity per unit band received by an omnidirectional hydrophone located at depth h, is given by
\begin{equation}
J_{o}(ah) = 2\pi J_{\infty}\int_0^{\pi/2} \cos^{n-1}\theta e^{-ah\sec\theta} \sin\theta d\theta
\label{eq:shorteq}
\end{equation}
where $J_{\infty}$ is the amplitude of the average intensity per unit band per unit solid angle radiated by a unit surface area, $\theta$ is the angle of the ray arriving at the hydrophone measured from the upward vertical direction, and $a = a(f)$ is related to the sound absorption coefficient $\alpha(f)$ by $\alpha h = -10\log(e^{-ah})$.
The index $n$ is 1 (2) for surface monopole (dipole) sources. Short calculates that the noise field below the horizontal plane of a hydrophone ($\theta > \pi$/2) at a height 91~m above the ocean bottom is significantly smaller than that above the horizontal and can be neglected. Therefore, Eq.~\ref{eq:shorteq} is integrated up to $\theta = \pi/2$ only taking into account direct paths from sources. Other assumptions made in deriving Eq.~\ref{eq:shorteq} are that the noise at the hydrophone is the incoherent sum of all intensities arriving, after attenuation, from the surface sources and that straight ray propagation is adequate \cite{Short}.   Thus, the depth dependent correction to the Knudsen spectra can be determined by adding the following expression to Eq.~(\ref{eq:seastate}).

\begin{equation}
10\log[J_{o}(ah)/J_{o}(0)]
\label{eq:shortoffset}
\end{equation}
In evaluating this frequency dependent offset, the average depth of the 49 hydrophones, $h = 1631$~m, is used.   The sound absorption in sea water is parameterized according to Fisher and Simmons \cite{Fisher} to evaluate $\alpha(f)$ and therefore $a(f)$. A temperature profile taken at TOTO every 7.6~m down to 1830~m is used in this evaluation. The resulting offset is always negative, as expected for an attenuation, and is applied to Eq.~(\ref{eq:seastate}). The result is also plotted in Fig.~\ref{fig:full_bandwidth_spectrum}.

An attempt to further improve the theoretical curve is made by considering the frequency-dependent directionality of the hydrophones above 15~kHz. This is obtained by modifying Eq.~\ref{eq:shorteq} into
\begin{equation} J_{eff}(ah) = 2\pi J_{\infty}\int_0^{\pi/2} \cos^{n-1}\theta e^{-ah\sec\theta} g(\theta,f) \sin\theta d\theta
\label{eq:modshort}
\end{equation} 
with the response function $g(\theta,f)$ provided as a look-up table by the hydrophone specifications. Eq.~\ref{eq:shortoffset} is then evaluated numerically and plotted in Fig.~\ref{fig:full_bandwidth_spectrum}.  It should be noted that the $P_{ss}$ fit to the data in both of the modified Knudsen spectra is performed after the Eq.~\ref{eq:shortoffset} correction and therefore, the numerical values of $P_{ss}$ are slightly different in all 3 curves. Also noteworthy is the fact that although using $n=1$ and $n=2$ create very small differences as also shown by Short\cite{Short}, the first case (monopole) produces a slightly better fit to data and is, therefore, used in the figure. However, the quality of data is not sufficient to significantly distinguish between the two models.

Clearly the curves including attenuation provide better descriptions of the data at high frequency.   Above 25~kHz the data appear to flatten faster than the model.   It is unclear as to whether this should be considered a real underwater effect or simply due to difficulties in modeling the response of the hydrophones.    It should be pointed out that the levels are too high to be considered as onset of thermal molecular noise, also shown in Fig.~\ref{fig:full_bandwidth_spectrum}.

\section{CONCLUSION}
The analysis of a large acoustic noise dataset collected at depths in excess of 1500~m and frequencies up to 40~kHz confirms that, even at substantial depths, a uniform surface distribution of sources alone is sufficient to describe ambient noise characteristics well up to 25~kHz.  A correlation observed between wind speed and overall noise levels at these depths during steady wind conditions further confirms this finding.  A description of the spectral shape above 15~kHz is found to require, at the depths considered here, proper account of the sound attenuation in water. This depth effect must therefore be included when defining parameters in acoustic neutrino detection, such as the transfer function and the signal to noise ratio, that require the expected noise spectrum.  

\begin{acknowledgments}
We would like to thank the US Navy and, in particular, D.~Deveau and T.~Kelly-Bissonnette for the hospitality and help at the AUTEC site. We are grateful to M.~Buckingham (Scripps Institution of Oceanography) and D.~Kapolka (Naval Postgraduate School) for sharing their insights and expertise. We also wish to acknowledge J.~Vandenbroucke (University of California, Berkeley), S.~Waldman (California Institute of Technology), and G. Walther (Stanford University) for illuminating discussions. This work was supported by NSF grant PHY-0457273.
\end{acknowledgments}

\bibliography{Kurahashi}% Produces the bibliography via BibTeX.

\begin{thebibliography}{12}
\expandafter\ifx\csname natexlab\endcsname\relax\def\natexlab#1{#1}\fi
\expandafter\ifx\csname bibnamefont\endcsname\relax
  \def\bibnamefont#1{#1}\fi
\expandafter\ifx\csname bibfnamefont\endcsname\relax
  \def\bibfnamefont#1{#1}\fi
\expandafter\ifx\csname citenamefont\endcsname\relax
  \def\citenamefont#1{#1}\fi
\expandafter\ifx\csname url\endcsname\relax
  \def\url#1{\texttt{#1}}\fi
\expandafter\ifx\csname urlprefix\endcsname\relax\def\urlprefix{URL }\fi
\providecommand{\bibinfo}[2]{#2}
\providecommand{\eprint}[2][]{\url{#2}}

\bibitem[{\citenamefont{Askar'yan}(1957)}]{Askaryan}
\bibinfo{author}{\bibfnamefont{G.~A.} \bibnamefont{Askar'yan}},
  \bibinfo{journal}{Soviet\ J.\ At.\ Energy} \textbf{\bibinfo{volume}{3}},
  \bibinfo{pages}{921} (\bibinfo{year}{1957}).

\bibitem[{\citenamefont{Fisher and Simmons}(1977)}]{Fisher}
\bibinfo{author}{\bibfnamefont{F.~H.} \bibnamefont{Fisher}} \bibnamefont{and}
  \bibinfo{author}{\bibfnamefont{V.~P.} \bibnamefont{Simmons}},
  \bibinfo{journal}{J.\ Acoust.\ Soc.\ Am.} \textbf{\bibinfo{volume}{62}},
  \bibinfo{pages}{558} (\bibinfo{year}{1977}).

\bibitem[{\citenamefont{Learned}(1979)}]{LearnedTheory}
\bibinfo{author}{\bibfnamefont{J.~G.} \bibnamefont{Learned}},
  \bibinfo{journal}{Phys.\ Rev.\ D} \textbf{\bibinfo{volume}{19}},
  \bibinfo{pages}{3293} (\bibinfo{year}{1979}).

\bibitem[{\citenamefont{Sulak et~al.}(1979)\citenamefont{Sulak, Armstrong,
  Baranger, Bregman, Levi, Mael, Strait, Bowen, Pifer, Polakos
  et~al.}}]{Learned}
\bibinfo{author}{\bibfnamefont{L.}~\bibnamefont{Sulak}},
  \bibinfo{author}{\bibfnamefont{T.}~\bibnamefont{Armstrong}},
  \bibinfo{author}{\bibfnamefont{H.}~\bibnamefont{Baranger}},
  \bibinfo{author}{\bibfnamefont{M.}~\bibnamefont{Bregman}},
  \bibinfo{author}{\bibfnamefont{M.}~\bibnamefont{Levi}},
  \bibinfo{author}{\bibfnamefont{D.}~\bibnamefont{Mael}},
  \bibinfo{author}{\bibfnamefont{J.}~\bibnamefont{Strait}},
  \bibinfo{author}{\bibfnamefont{T.}~\bibnamefont{Bowen}},
  \bibinfo{author}{\bibfnamefont{A.~E.} \bibnamefont{Pifer}},
  \bibinfo{author}{\bibfnamefont{P.~A.} \bibnamefont{Polakos}},
  \bibnamefont{et~al.}, \bibinfo{journal}{Nucl.\ Instrum.\ Methods}
  \textbf{\bibinfo{volume}{161}}, \bibinfo{pages}{203} (\bibinfo{year}{1979}).

\bibitem[{\citenamefont{Urick}(1983)}]{Urick}
\bibinfo{author}{\bibfnamefont{R.~J.} \bibnamefont{Urick}},
  \emph{\bibinfo{title}{Principles of Underwater Sound}}
  (\bibinfo{publisher}{McGraw-Hill}, \bibinfo{year}{1983}).

\bibitem[{\citenamefont{Knudsen et~al.}(1948)\citenamefont{Knudsen, Alford, and
  Emling}}]{Knudsen}
\bibinfo{author}{\bibfnamefont{V.~O.} \bibnamefont{Knudsen}},
  \bibinfo{author}{\bibfnamefont{R.~S.} \bibnamefont{Alford}},
  \bibnamefont{and} \bibinfo{author}{\bibfnamefont{J.~W.}
  \bibnamefont{Emling}}, \bibinfo{journal}{J.\ Mar.\ Res}
  \textbf{\bibinfo{volume}{3}}, \bibinfo{pages}{410} (\bibinfo{year}{1948}).

\bibitem[{\citenamefont{Marsh}(1963)}]{Marsh}
\bibinfo{author}{\bibfnamefont{H.~W.} \bibnamefont{Marsh}},
  \bibinfo{journal}{J.\ Acoust.\ Soc.\ Am.} \textbf{\bibinfo{volume}{35}},
  \bibinfo{pages}{409} (\bibinfo{year}{1963}).

\bibitem[{\citenamefont{Short}(2005)}]{Short}
\bibinfo{author}{\bibfnamefont{J.}~\bibnamefont{Short}}, \bibinfo{journal}{IEEE
  J.\ Ocean.\ Eng.} \textbf{\bibinfo{volume}{30}}, \bibinfo{pages}{267}
  (\bibinfo{year}{2005}).

\bibitem[{\citenamefont{Lehtinen et~al.}(2002)\citenamefont{Lehtinen, Adam,
  Gratta, Berger, and Buckingham}}]{Lehtinen}
\bibinfo{author}{\bibfnamefont{N.~G.} \bibnamefont{Lehtinen}},
  \bibinfo{author}{\bibfnamefont{S.}~\bibnamefont{Adam}},
  \bibinfo{author}{\bibfnamefont{G.}~\bibnamefont{Gratta}},
  \bibinfo{author}{\bibfnamefont{T.~K.} \bibnamefont{Berger}},
  \bibnamefont{and} \bibinfo{author}{\bibfnamefont{M.~J.}
  \bibnamefont{Buckingham}}, \bibinfo{journal}{Astropart.\ Phys.}
  \textbf{\bibinfo{volume}{17}}, \bibinfo{pages}{279} (\bibinfo{year}{2002}).

\bibitem[{\citenamefont{Vandenbroucke et~al.}(2004)\citenamefont{Vandenbroucke,
  Gratta, and Lehtinen}}]{Vandenbroucke}
\bibinfo{author}{\bibfnamefont{J.}~\bibnamefont{Vandenbroucke}},
  \bibinfo{author}{\bibfnamefont{G.}~\bibnamefont{Gratta}}, \bibnamefont{and}
  \bibinfo{author}{\bibfnamefont{N.}~\bibnamefont{Lehtinen}},
  \bibinfo{journal}{ApJ} \textbf{\bibinfo{volume}{621}}, \bibinfo{pages}{301}
  (\bibinfo{year}{2004}).

\bibitem[{\citenamefont{Wenz}(1962)}]{Wenz}
\bibinfo{author}{\bibfnamefont{G.~M.} \bibnamefont{Wenz}},
  \bibinfo{journal}{J.\ Acoust.\ Soc.\ Am.} \textbf{\bibinfo{volume}{34}},
  \bibinfo{pages}{1936} (\bibinfo{year}{1962}).

\bibitem[{\citenamefont{Everest et~al.}(1948)\citenamefont{Everest, Young, and
  Johnson}}]{Everest}
\bibinfo{author}{\bibfnamefont{F.~A.} \bibnamefont{Everest}},
  \bibinfo{author}{\bibfnamefont{R.~W.} \bibnamefont{Young}}, \bibnamefont{and}
  \bibinfo{author}{\bibfnamefont{M.~W.} \bibnamefont{Johnson}},
  \bibinfo{journal}{J.\ Acoust.\ Soc.\ Am.} \textbf{\bibinfo{volume}{20}},
  \bibinfo{pages}{137} (\bibinfo{year}{1948}).

\end{thebibliography}

\end{document}